\documentstyle[amssymb,aps,prb,twocolumn]{revtex}

\begin{document}
\title{Ballistic versus diffusive magnetoresistance of a magnetic point contact}
\author{L.R. Tagirov$^{1}$, B.P. Vodopyanov$^{2}$, and K.B. Efetov$^{3,4}$}
\address{$^{1}$Kazan State University, Kazan 420008, Russia\\
$^2$Kazan Physico-Technical Institute of RAS, Kazan 420029, Russia\\
$^3$Theoretische Physik III, Ruhr-Universit\"at Bochum, 44780 Bochum, Germany
\\
$^4$L.D. Landau Institute for Theoretical Physics, Moscow, Russia}
\date{\today{}}
\maketitle
\draft

\begin{abstract}
The quasiclassical theory of a nanosize point contacts (PC) between two
ferromagnets is developed. The maximum available magnetoresistance values in
PC are calculated for ballistic {\it versus} diffusive transport through the
area of a contact. In the ballistic regime the magnetoresistance in excess
of few hundreds percents is obtained for the iron-group ferromagnets. The
necessary conditions for realization of so large magnetoresistance in PC,
and the experimental results by Garc\'\i a {\it et al} are discussed.
\end{abstract}

\pacs{PACS numbers: 74.80.Dm, 74.50.+r, 74.62.-c}

In recent experiments on study of Ni-Ni and Co-Co point contacts (PC), a
surprisingly high negative magnetoresistance exceeding $200\%$ has been
discovered.\cite{Garcia1,Garcia2} The set up of the experiment was typical
for observation of giant magnetoresistance (GMR), the effect observed
earlier in hybrid systems involving ferromagnetic and normal multilayer
metals.\cite{Gijs97,Anserm} However, for the multilayer structures the
typical change of the resistance reached $10\%-30\%,$ which is considerably
lower than the corresponding values of Refs. \onlinecite{Garcia1,Garcia2}.
So, one can come easily to the conclusion that the main contribution to the
MR comes from the region of the PC.

A negative magnetoresistance can be due to scattering on domain walls (DW)
and this effect has been considered in a number of works \cite
{Falicov,Berger,Levy,VanHoof} giving typical values of MR in a range of few
percents. Such considerably low values of the MR were obtained assuming that
realistic widths of the DW were large, which resulted in a low scattering
amplitudes. Considering sharp DW in the ballistic regime one comes to values
$\sim 70\%$.\cite{VanHoof}

The fact that a sharp DW may give large MR was used in Ref.
\onlinecite{Garcia2} to explain the anomalously large values of MR in the
experiments on the point contacts.\cite{Garcia1,Garcia2} However, the
theory in Ref. 2 is the perturbation theory, it can not be applied
to the explanation of 300\% effect.
The diminishing of
the width of DW, when decreasing the size of the constriction was
demonstrated by Bruno.\cite{Bruno} The DW width becomes comparable with PC
length, and magnetization rotates almost abruptly inside the constriction.
This conclusion holds until the diameter of PC is smaller than its actual
length. With further increase of constriction size (diameter) the wall will
bend outside of PC, and simple energy considerations show that the DW width
will be of the order of PC size.\cite{Volkov}

The regime of conductance quantization in magnetic PC has been considered by
Imamura {\it et al} .\cite{Imamura} They obtained that, if the spin of
conduction electron cannot rotate in DW pinned to the constriction, then
magnetoresistance acquires oscillations as a function of PC size with
amplitude exceeding 1000\%.

In this paper we develop a quasiclassical theory of electric transport
through magnetic PC taking into account scattering by impurities, thus
covering the ballistic $\left( l>a\right) $ and diffusive $\left( l<a\right)
$ regimes ($l$ is the mean free path and $a$ is the radius of the contact).
The typical PC size, which is beyond the quantization regime, $2a\geq 8$\AA
, may be well described within the quasiclassical (QC) approximation ($2a\gg
\lambda _F=2\pi /p_F\sim 6$\AA , $\lambda _F$ and $p_F$ are the Fermi
wavelength and momentum).

We believe that extremely large magnetoresistance can be obtained, if the
strong reflection of spin-polarized current carriers on the PC area is
achieved at {\em antiparallel} (AP) alignment of magnetizations in
contacting ferromagnets. This is realized, if there is mismatch in the
spin-subbands Fermi-momenta of contacting magnets. For AP alignment $
p_{F1\uparrow }=p_{F2\downarrow }$, and $p_{F1\downarrow }=p_{F2\uparrow }$.
Let us assume that $p_{F1\uparrow }\gg p_{F1\downarrow }$. Then subband with
the smaller value of the Fermi momenta, which is minority subband, can not
accept momenta transferred from the opposite side of the PC, which is
majority subband with the same spin projection. As a result, only a narrow
incidence angles cone around the normal to the interface is responsible for
the charge transport across the PC. Electrons with more inclined
trajectories are completely reflected. Thus, the partial transmission at the
steep incidence, and the total reflection at slanting incidence provide high
boundary resistance of PC.

The necessary condition for realization of the above scenario is the
conservation of electron spin orientation when crossing the domain wall. The
orientation conserves, if the DW width $d_w$ is shorter than the length $d_s$
, at which the electron spin quantization axis adjusts the varying direction
of local exchange field. For ballistic transmission through PC $d_s=v_FT_1$,
where $T_1$ is the longitudinal relaxation time of conduction electron
magnetization - the Overhauser time.\cite{Overhauser} At this condition the
transmission process looks like transmission through abrupt DW, and the
description of the electron transport through PC with boundary conditions at
PC interface is valid.

The PC model we consider is the circular hole of the radius $a$ made in a
membrane, which divides the space on two half-spaces, occupied by
single-domain ferromagnetic metals. The membrane is impenetrable for the
quasiparticles carrying a current, however the thickness of the membrane in
the model is assumed to be vanishing. The $z-$axis of coordinate system is
chosen perpendicular to the membrane plane. The electron motion on both
sides of the contact can be described by the equations for quasiclassical
(QC) Green functions derived by Zaitsev.\cite{Zaits1} They are in fact the
Boltzmann equations in the $\tau -$approximation:
\begin{eqnarray}
v_z\frac{\partial g_a}{\partial z}+{\bf v}_{\parallel }\frac{\partial g_s}{
\partial \overrightarrow{\rho }}+\frac 1\tau \left( g_s-\overline{g}
_s\right) &=&0,  \nonumber  \\
v_z\frac{\partial g_s}{\partial z}+{\bf v}_{\parallel }\frac{\partial g_a}{
\partial \overrightarrow{\rho }}+\frac{g_a}\tau &=&0.  \label{eq1}
\end{eqnarray}
$g_s$ and $g_a$ are symmetric and antisymmetric with respect to $z-$
projection of quasiparticle momentum QC GF (Green functions integrated over
the energy variable), ${\bf v}$ is the vector of the Fermi velocity, $
v_z=v_F\cos \theta $, $v_{\parallel }^2=v_F^2-v_z^2$ , angle $\theta $ is
measured from the $z-$axis, $v_F$ is the modulus of ${\bf v}$, the bar over $
\overline{g}_c$ means the averaging over the solid angle. We assume that the
spin-mixing process is weak, therefore we consider spin channels as
independent and omit the spin-channel indices in (\ref{eq1}) and expressions
below.

The boundary conditions to equation (\ref{eq1}) for the specular scattering (
$p_{F1\alpha }\sin \theta _1=p_{F2\alpha }\sin \theta _2\equiv p_{\Vert }$)
at the interface $z=0$ are:\cite{Zaits1}
\begin{eqnarray}
g_{a1}(0) &=&g_{a2}(0)=\left\{
{\displaystyle {g_a(0),\,\,\,\,p_{\Vert }<p_{F1},p_{F2} \atop 0,\,
\,\,\,\min (p_{F1},p_{F2})<p_{\Vert }}}
\right. ,  \nonumber \\
2Rg_a(0) &=&-D\left( g_{s2}-g_{s1}\right) ,  \label{eq2}
\end{eqnarray}
where subscript 1 or 2 labels left- or right-hand side of the contact,
respectively, $p_{Fi}$ is the Fermi momentum of $i$-th side, $p_{\Vert }$ is
the projection of the Fermi momentum vector on the PC plane. $D$ and $R=1-D$
are the exact quantum mechanical transmission and reflection coefficients
that can be considered either as phenomenological parameters or calculated
for models of interest. The second line in the first boundary condition in (
\ref{eq2}) explicitly quantifies the total reflection for inclined
trajectories, described qualitatively above.

The density of a current through the contact may be written as
\begin{equation}
j^z(z,\overrightarrow{\rho },t)=-\frac{ep_{F\min }^2}{2\pi }{
\int\limits_0^{\pi /2}}d\Omega _\theta \cos \theta g_a\left( z,
\overrightarrow{\rho },t\right) .  \label{eq3}
\end{equation}
The total current through the area of the contact is
\begin{equation}
I^z(z\rightarrow 0,t)=a{\int\limits_0^\infty }dkJ_1(ka)j^z(0,k,t).
\label{eq4}
\end{equation}
In the above equations $p_{F\min }=\min \left( p_{F1},p_{F2}\right) $, $
J_1(x)$ is the Bessel functions, $j^z(0,k,t)$ is the Fourier-transform of
current density, Eq. (\ref{eq3}), over the in-plane coordinate ${\bf \rho }$
. The cylindrical symmetry of the problem has been used upon derivation of
Eq. (\ref{eq4}).

We search a solution for $g_s$ in the form ($k_B=\hbar =1$):

\begin{equation}
g_s(\varepsilon )=\tanh \frac \varepsilon {2T}+f_s(\varepsilon ),
\label{eq5}
\end{equation}
where the first term is the equilibrium value of $g_s$ in the leads far away
of PC. Substitution of (\ref{eq5}) into (\ref{eq1}) and Fourier
transformation over the variable ${\bf \rho }$ leads to equations, the exact
solution of which reads
\begin{equation}
f_s(z)=g_a(z){\rm sgn}(z)+\frac 1{l_z}{\int\limits_{-\infty }^\infty }d\xi
e^{-\varkappa \left| \xi -z\right| }\overline{f}_s(\xi ,k),  \label{eq15}
\end{equation}
where
\begin{equation}
\varkappa =\frac{1-i{\bf kl}_{\parallel }}{l_z},  \label{eq15-1}
\end{equation}
$l=\tau v_F$ is the mean free path, $l_z=l\cos \theta $, $l_{\parallel
}^2=l^2-l_z^2$. Integrating Eq. (\ref{eq15}) over the solid angle we obtain
\begin{equation}
\overline{f}_s(z>0)=\overline{g}_a+{\int\limits_z^\infty }d\xi K(\xi -z)
\overline{f}_s(\xi ,k),  \label{eq16}
\end{equation}
where the kernel $K(\eta )$ is ($x=\cos \theta $)
\begin{equation}
K(\eta )=\frac 1l{\int\limits_0^1}dx\frac{e^{-\frac \eta {lx}}}xJ_0(k\eta
\frac{\sqrt{1-x^2}}x).  \label{eq17}
\end{equation}
If the mean free path $l$ is short ($l\ll a$), the second term in Eqs. (\ref
{eq15}) and (\ref{eq16}) dominates and the integrand of Eq. (\ref{eq16}) is
the product of rapidly decreasing on the distance $l$ kernel $K(\eta )$ and
slowly decreasing function $\overline{f}_s$. That is why we may take out $
\overline{f}_c(k,\xi )$ from the integral (\ref{eq16}) at the point $\xi =z.$
Within this approximation we obtain
\begin{equation}
\bar f_s\left( z,k\right) =\bar g_a\left( z,k\right) \left( 1-\lambda \left(
k\right) \right) ^{-1}  \label{eq18}
\end{equation}
where
\begin{equation}
\lambda (k)={\int\limits_0^\infty }d\xi K(\xi -z)=\frac 1{kl}\arctan kl.
\label{eq19}
\end{equation}
Substituting Eq. (\ref{eq18}) into Eq. (\ref{eq15}), and using the boundary
conditions (\ref{eq2}) we obtain the equation for the antisymmetric
combination $g_a$:
\begin{eqnarray}
g_a(0,k) &=&-\frac 12D\left( \tanh \frac \varepsilon {2T}-\tanh \frac{
\varepsilon -eV}{2T}\right) \gamma _k  \nonumber \\
&&\ \ \ \ \ \ \ \ \ \ \ -\frac D{1-\lambda _1}\frac 1{2l_{z1}}{
\int\limits_{-\infty }^0}d\xi e^{\varkappa _1\xi }\overline{g}_{a1}(\xi )
\nonumber \\
&&\ \ \ \,\,\,\,\,\,\,\,\,\,\,\,\,\,-\frac D{1-\lambda _2}\frac 1{2l_{z2}}{
\int\limits_0^\infty }d\xi e^{-\varkappa _2\xi }\overline{g}_{a2}(\xi ),
\label{eq22}
\end{eqnarray}
where
\begin{equation}
\gamma _k={\int\limits_0^a}\rho d\rho {\int\limits_0^{2\pi }}e^{i{\bf k}
\overrightarrow{{\bf \rho }}}d\varphi =\frac{2\pi a}kJ_1(ka),  \label{eq23}
\end{equation}
$V$ is the bias voltage.

To find $g_a(0,k)$ we average Eq. (\ref{eq22}) over the solid angle, exploit
the continuity of $g_a$ at the interface, Eq. (\ref{eq2}), and, again, use
the fact that in the limit $l\ll a$ the kernel in the integral over $x$ in
the second and third terms of averaged Eq. (\ref{eq22}) is a function,
rapidly decreasing at distance $l$. Of course, in the ballistic regime ($l>a$
) this approximation is no longer valid, but, in this regime the first
(exact) term in Eq. (\ref{eq22}) dominates the approximate terms with
integrals. So, in the ballistic limit the approximation done does not bring
a big error either. Although the approximation may not be valid in the
intermediate regime, the suggested scheme can be used as an interpolation.

Now we find easily $\overline{g}_a(0,k)$, make consecutive substitutions
into (\ref{eq22}), (\ref{eq3}), (\ref{eq4}), and, finally, obtain the
general expression for the current through PC:
\begin{equation}
I^z=\frac{e^2p_{F\min }^2a^2V}{2\pi }{\int\limits_0^\infty }\frac{dk}k
J_1^2(ka)\left\langle D\,F(k,\theta )\cos \theta \right\rangle ,
\label{eq28}
\end{equation}
where
\begin{eqnarray}
F(k,\theta ) &=&1-\left[ \frac 1{2\left( 1-\lambda _1\right) \varkappa
_1l_{z1}}+\frac 1{2\left( 1-\lambda _2\right) \varkappa _2l_{z2}}\right]
\nonumber \\
&&\ \,\,\,\,\,\ \ \ \times \frac{\overline{D}}{1+\frac{\widetilde{\lambda }_1
}{2\left( 1-\lambda _1\right) }+\frac{\widetilde{\lambda }_2}{2\left(
1-\lambda _2\right) }},  \label{eq29}
\end{eqnarray}

\begin{equation}
\widetilde{\lambda }_i=\left\langle \frac D{\varkappa _il_{zi}}
\,\right\rangle ={\int\limits_0^1}dx\frac{D(x)}{\sqrt{1+k^2l_i^2(1-x^2)}},
\label{eq30}
\end{equation}
$<\ldots >$ means averaging over the solid angle. Eqs. (\ref{eq28}) and (\ref
{eq29}) are the basic analytical result of the paper, which expresses the
current in terms of parameters $D$, $l$, $a$, $p_F$ characterizing the
system.

Now we calculate the magnetoresistance of PC between two identical
ferromagnets. It can be expressed via the conductances $\sigma =I/V$ as
follows
\begin{equation}
MR=\frac{R^{AP}-R^P}{R^P}=\frac{\sigma ^P-\sigma ^{AP}}{\sigma ^{AP}},
\label{eq36}
\end{equation}
where $R^P$ $(\sigma ^P)$ stands for the resistance (conductance) at {\em
parallel} alignment of magnetizations of contacting ferromagnets, and $R^{AP}
$ $(\sigma ^{AP})$ is for the {\em antiparallel} alignment of
magnetizations. For the {\em parallel} alignment the net current is the sum
of currents for both (independent) spin channels, $D=1$, $\widetilde{\lambda
}_i=\lambda _i$. Labelling the quantities by arrow-up/down notations we
write down
\begin{eqnarray}
\sigma ^P &=&\sigma _{\uparrow \uparrow }^z+\sigma _{\downarrow \downarrow
}^z=\frac{e^2\left( p_{F\uparrow }^2+p_{F\downarrow }^2\right) \left( \pi
a^2\right) }{4\pi ^2}{\int\limits_0^\infty }\frac{dk}kJ_1^2(ka)  \nonumber \\
&&\ \times \left\{ \frac{p_{F\uparrow }^2}{p_{F\uparrow }^2+p_{F\downarrow
}^2}\frac{k^2l_{\uparrow }^2}{\left( 1+\sqrt{1+k^2l_{\uparrow }^2}\right) ^2}
+(\uparrow \rightleftarrows \downarrow )\right\} .  \label{eq37}
\end{eqnarray}
The prefactor in Eq. (\ref{eq37}) is nothing but the sum of Sharvin \cite
{Sharvin} conductances for the spin channels. For the AP alignment of
magnetizations the conductance is
\[
\sigma ^{AP}=\frac{e^2p_{F\downarrow }^2\left( \pi a^2\right) }{\pi ^2}{
\int\limits_0^\infty }\frac{dk}kJ_1^2(ka){\int\limits_0^1}dxx\left(
D(x)\right) _{\uparrow \downarrow }
\]
\begin{eqnarray}
&&\ \ \ \ \times \left\{ 1-\left[ \frac{1-\lambda ^{\uparrow }}{\sqrt{
1+k^2l_{\uparrow }^2\left( 1-x^2\right) }}+\frac{1-\lambda ^{\downarrow }}{
\sqrt{1+k^2l_{\downarrow }^2\left( 1-x^2\right) }}\right] \right.   \nonumber
\\
&&\ \ \ \ \times \left. \frac{\left( \overline{D}\right) _{\uparrow
\downarrow }}{2\left( 1-\lambda ^{\uparrow }\right) \left( 1-\lambda
^{\downarrow }\right) +\widetilde{\lambda }_{\uparrow \downarrow }^{\uparrow
}(1-\lambda ^{\downarrow })+\widetilde{\lambda }_{\uparrow \downarrow
}^{\downarrow }(1-\lambda ^{\uparrow })}\right\} ,  \label{eq38}
\end{eqnarray}
where $\left( D(x)\right) _{\uparrow \downarrow }$ stands for the
transmission coefficient of the interface at AP alignment. For the mechanism
of magnetoresistance discussed above, $\left( D(x)\right) _{\uparrow
\downarrow }$ can be found from the solution of Schr\"odinger equation for
the particle moving in the step-like potential landscape \cite{LL}

\begin{equation}
\left( D(x)\right) _{\uparrow \downarrow }=\frac{4(v_{z1}^{\uparrow
})_{\uparrow }(v_{z2}^{\uparrow })_{\downarrow }}{\left( (v_{z1}^{\uparrow
})_{\uparrow }+(v_{z2}^{\uparrow })_{\downarrow }\right) ^2}=\left(
D(x)\right) _{\downarrow \uparrow }  \label{eq39}
\end{equation}
with $v_{z2}^{\uparrow }=v_{z1}^{\downarrow }$ for the {\em antiparallel}
alignment. The transmission coefficient (\ref{eq39}) gives maximum available
magnetoresistance values for a particular parameters choice. Neglecting the
difference of the effective masses in the spin-subbands we may write down
\begin{equation}
\left( D(x)\right) _{\uparrow \downarrow }\simeq \frac{4x\sqrt{b^2+x^2}}{
\left( x+\sqrt{b^2+x^2}\right) ^2},  \label{eq40}
\end{equation}
where
\begin{equation}
b^2=\frac{1-\delta ^2}{\delta ^2},\,\,\,\,\delta =\frac{p_{F\downarrow }}{
p_{F\uparrow }}=\frac{v_{F\downarrow }}{v_{F\uparrow }}\leq 1.  \label{eq41}
\end{equation}
For the purely ballistic transport ($a/l_{\uparrow }\rightarrow 0$, where $
l_{\uparrow }$ ($l_{\downarrow }$) is the majority (minority) electrons mean
free path) all integrals in Eqs. (\ref{eq37}), (\ref{eq38}) are evaluated
analytically, and magnetoresistance reads
\begin{equation}
MR=\frac{(1-\delta )\left\{ 5\delta ^3+15\delta ^2+9\delta +3\right\} }{
8\delta ^3(\delta +2)}.  \label{eq47}
\end{equation}
If $\delta =1$, then $MR=0$, {\it i.e.} the magnetoresistance vanishes. For
the set of $\delta $ values we obtain from (\ref{eq47}): $\delta =0.5$, $
MR=238\%$; $\delta =0.4$, $MR=455\%$; $\delta =0.33$, $MR=780\%$; $\delta
=0.3$, $MR=1012\%$.

In general case the angular integrals in (\ref{eq30}) and (\ref{eq38}) can
be still evaluated analytically, whereas the integrations over $k$ can be
done only numerically. The results for the magnetoresistance (\ref{eq36}) as
function of the contact radius are shown on Fig. 1. The curves show the
maximum available MR, that could be realized in PC with physical parameters
displayed on the figure. MR exponentially drops when the size of the contact
approaches the mean free path of a material. Then it shows a smooth
crossover from ballistic to diffusive regimes of conduction.

Let us discuss the experimental data on magnetoresistance of magnetic PC by
Garc\'\i a {\it et al.} Ni-Ni PC showed maximal $MR\simeq 280\%$,\cite
{Garcia1} Co-Co PC showed maximal $MR\simeq 230\%$.\cite{Garcia2} In the
recent paper \cite{Garcia3} they quote maximal $MR\simeq 33\%$ for Fe-Fe PC.
To obtain the $MR$ values 280\% (Ni) and 230\% (Co) we have to use the
values $\delta ($Ni$)\simeq 0.47$ and $\delta ($Co$)\simeq 0.5$. These
numbers are in the range of the values, obtained experimentally from the
single photon threshold photoemission, $\delta ($Co$)\simeq 0.4$, \cite
{Grobli} and from ferromagnet/superconductor point contact spectroscopy: $
\delta ($Ni$)\simeq 0.59-0.65$,\cite{Soulen} $\delta ($Ni$)\simeq 0.71$;\cite
{Buhrman} $\delta ($Co$)\simeq 0.62-0.65$,\cite{Soulen} $\delta ($Co$)\simeq
0.68$.\cite{Buhrman}

If we use the experimental data of Ref. \onlinecite{Soulen} for iron, $
\delta ($Fe$)\simeq 0.59-0.65$, then in our theory we obtain $MR($Fe$
)=(100-140)\%$, which is larger than the experimentally measured 33\%.\cite
{Garcia3} The justification of our model suggests that observed MR does not
solely confined to a value of polarization $\delta $. We believe that the
basic condition for observation of upper MR limit, $d_w\ll d_s$, is not
fulfilled in the Fe-Fe PC experiment.\cite{Garcia3} $T_1$ is proportional to
the squared magnetic moment and the integral of exchange between conduction
electrons and localized moments, and proportional to conduction electrons
density of states at Fermi level. All these physical parameters for iron are
larger than for cobalt, and especially than for nickel. Therefore we expect,
that $T_1$(Fe) at least one order of magnitude shorter than $T_1$ for Co and
Ni. When $d_s$(Fe$)\sim d_w$(Fe), the electron spin almost tracks the local
exchange field in the domain wall. As a result the reflection of the
electrons from DW decreases, and the observed MR does not reach its maximal
value.

Let us discuss now the magnetoresistance in the diffusive regime of
transport, when the radius of nanohole is much larger than the
mean free path ($a\gg l_{\uparrow },l_{\downarrow }$).
The giant MR values can be
obtained, if the condition of validity of our model, $d_w\ll d_s$, will be
realized in an experiment. In the opposite limit, $d_w>d_s$, when PC size is
so large that DW becomes smooth and wide, the electron spin will track the
local exchange field in the domain wall, and MR will level off at Levy-Zhang
\cite{Levy} impurity scattering enhancement mechanism, which can give 2-11\%
magnetoresistance. The requirement of abrupt DW with constant width,
irrespective of the PC size, can be technologically controlled, if very thin
(2-4 monolayers of the thickness $\sim \lambda _F$) nonmagnetic interlayer
is deposited on the PC plane before depositing the second electrode. Then,
just like in CPP transport in multilayers,\cite{Gijs97,Anserm} the
contacting domains will be exchange decoupled, so the magnetization will
acquire sudden reversal within the spacer thickness $\sim \lambda _F$. In
this case our analysis is valid for an {\em arbitrary} size of PC.

This work has been supported by Deutsche SFB 491. L.R.T. and B.P.V.
acknowledge the support by the Russian Science Foundation through the grant
N 00-02-16328 and by CRDF through the grant REC-007. We are grateful to
Profs P. Bruno, G.B. Teitelbaum, A.F. Volkov, K.Westerholt and H. Zabel for
discussions of the results.

\begin{center}
{\bf Figure captions}
\end{center}

Fig.1. The dependence of magnetoresistance on the PC radius.

\end{document}